\documentclass{icrc29}
\usepackage{graphicx,amssymb,amsmath,times}
\begin{document}
\title[Upper limit on primary photon fraction from the P.~Auger
Observatory]{Upper limit on the primary photon fraction from the
Pierre Auger Observatory}
\author[The Pierre Auger Collaboration]
{The Pierre Auger Collaboration}
\presenter{Presenter: M.~Risse (markus.risse@ik.fzk.de), \  
ger-risse-M-abs2-he14-oral}

\maketitle

\begin{abstract}

Based on observations of the depth of shower maximum performed with the
hybrid detector of the Auger Observatory,
an upper limit on the cosmic-ray photon fraction of 26\%
(at 95\% confidence level)
is derived for primary energies above $10^{19}$~eV.
Additional observables recorded with the surface detector
array, available for a sub-set of the data sample,
support the conclusion that a photon origin of the observed events is
not favoured.

\end{abstract}

\section{Introduction}

One of the key observables to distinguish between model predictions
on the origin of the highest-energy cosmic rays
is the fraction of primary cosmic-ray photons.
In non-acceleration
(``top-down'') models a significant fraction of the generated
particles are photons~\cite{models}.
Air showers initiated by photons at energies above $10^{19}$~eV
are in general expected to have a relatively large depth of shower
maximum $X_{\rm max}$ and
fewer secondary muons compared to nuclear primaries.
Previous upper limits on the photon fraction were derived from surface
array data of the Haverah Park and AGASA
experiments~\cite{ave,shinozaki,risse05}.

We report an analysis of data recorded by the
Auger Observatory~\cite{auger}.
The photon upper limit derived
here is based on the direct observation of the longitudinal air
shower profile and makes use of the hybrid detection technique:
$X_{\rm max}$ is used as discriminant observable.
The information from triggered surface detectors in hybrid
events considerably reduces the uncertainty in shower track geometry.

For a sub-set of the event sample used in this analysis, a variety
of surface detector observables
is available.
The additional discrimination power of these observables is demonstrated.

\section{Data}

The data are taken with a total of 12 fluorescence
telescopes~\cite{bellido}, situated at
two different telescope sites, during the period January 2004
to April 2005. The number of deployed surface detector (SD)
stations ~\cite{bertou05} grew
from $\sim$200 to $\sim$800 during this time.
For the analysis, hybrid events were selected, i.e.~showers observed
both by (at least one) surface tank and telescope~\cite{mostafa}.
Even for one triggered tank only, the additional timing constraint allows
a significantly improved geometry fit to the observed profile
which leads to a reduced uncertainty in the reconstructed $X_{\rm max}$.
The following criteria are applied for event selection:
\begin{itemize}
\item[$\bullet$]
to maximize the reconstruction quality:
geometry and profile fits succeeded,
$X_{\rm max}$ observed,
track length in field of view $>$400~g~cm$^{-2}$,
minimum viewing angle $>$18$^\circ$,
primary photon energy~~$\lg E/$eV$>$19.0;
\item[$\bullet$]
to achieve comparable detector acceptance to 
photon and nuclear primaries:
primary zenith angle $>$35$^\circ$,
distance of telescope to shower axis $<$24~km + $f(E)$, with
$f(E)= 12$~km$\cdot(\lg E/$eV$-19.0)$.
\end{itemize}
The reconstruction is based on an end-to-end calibration of the
fluorescence telescopes~\cite{brack}, on monitoring data of
local atmospheric conditions~\cite{keilhauer,roberts},
and includes an improved subtraction of Cherenkov 
light~\cite{nerling}
and reconstruction of energy deposit profiles for deriving the
primary energy.
The decreased fraction of missing energy in primary photon showers
is accounted for.
In total, 16 events with energies above $10^{19}$~eV  are selected.

The total uncertainty $\Delta X_{\rm max}^{\rm tot}$ of the
reconstructed depth of shower maximum is composed of
several contributions which, in general, vary from event to event.
A conservative estimate of the current $X_{\rm max}$ uncertainties
gives $\Delta X_{\rm max}^{\rm tot}\simeq$ 40~g~cm$^{-2}$.
Among the main contributions, each one in general well below
$\Delta X_{\rm max}=$15~g~cm$^{-2}$, are
the statistical uncertainty from the profile fit,
the uncertainty in shower geometry,
the uncertainty in atmospheric conditions such as the air
density profile, and
the uncertainty in the reconstructed primary energy, which is taken
as input for the primary photon simulation.

For each event, high-statistics shower simulations are performed for
photons for the specific event conditions.
Possible
cascading of photons in the geo\-magnetic field is simulated with
PRE\-SHO\-WER~\cite{homola}.
Shower development in air, including the LPM effect~\cite{lpm}, is
calculated with CORSIKA~\cite{heck}.
The Particle Data Group extrapolation of the photonuclear
cross-section~\cite{pdg} and QGSJET~01~\cite{qgsjet}
as hadron event generator are adopted.

A simulation study of the detector acceptance to photons and nuclear
primaries has been conducted.
For the chosen cuts, the ratio of the acceptance to photon-induced showers
to that of nuclear primaries (proton or iron nuclei) is $\epsilon = 0.88$.
A corresponding correction is applied to the derived photon limit.

\section{Results}

Fig.~1 shows as an example an event of 11~EeV primary energy
observed with
$X_{\rm max} = 744$~g~cm$^{-2}$, compared to the corresponding
$X_{\rm max}$ distribution expected for primary photons.
With $<$$X_{\rm max}^\gamma$$> = 1020$~g~cm$^{-2}$, photon showers are on
average expected to reach maximum at depths considerably greater than
observed.
Shower-to-shower fluctuations are large due to the LPM effect
(rms of 80~g~cm$^{-2}$) and well in excess of the measurement
uncertainty.
For all 16 events, the observed $X_{\rm max}$ is well below the average value
expected for photons.
The $X_{\rm max}$ distribution of the data is also displayed in Fig.~1.
\begin{figure}[b]
\noindent
\begin{minipage}[l]{.48\linewidth}
\includegraphics[width=.93\textwidth]{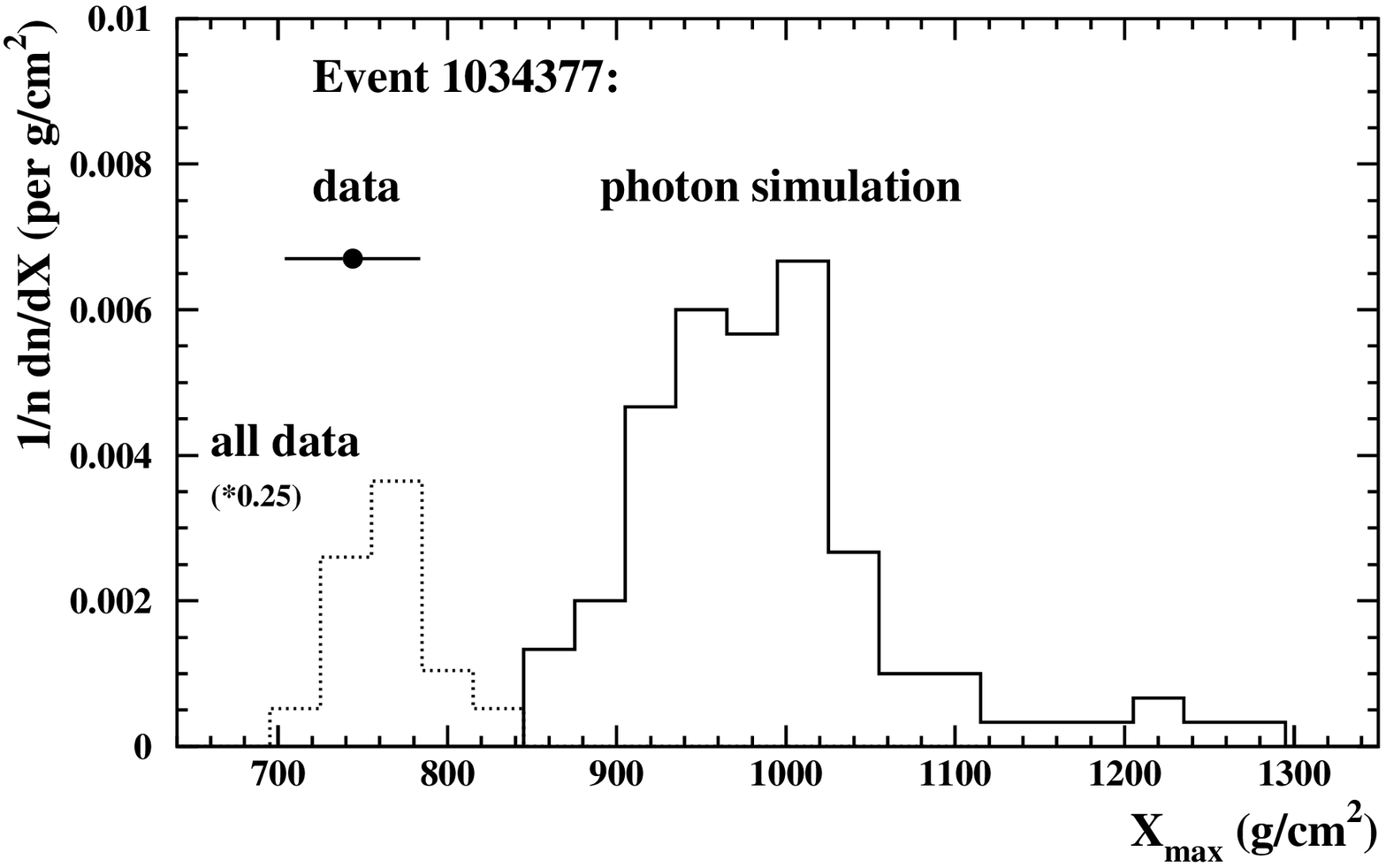}
\caption{Example of $X_{\rm max}$ measured in an individual shower
of 11~EeV (point
with error bar) compared to the $X_{\rm max}$ distribution
expected for photon showers (solid line).
Also shown the $X_{\rm max}$ distribution of the data sample (dashed line;
normalization changed as indicated).}
\label{fig1}
\end{minipage}\hfill
\begin{minipage}[c]{.48\linewidth}
\includegraphics[width=.98\textwidth]{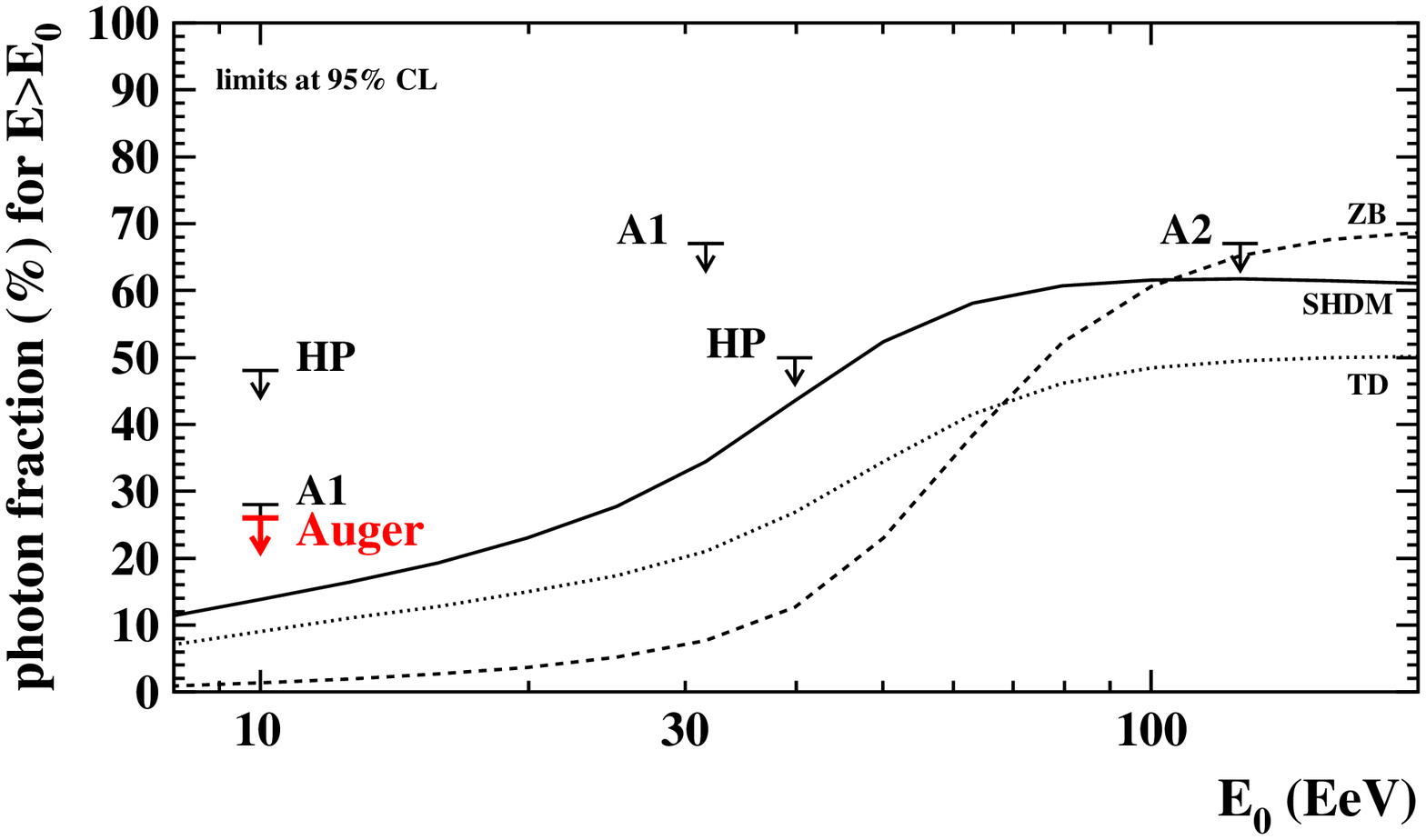}
\caption{
Upper limits (95\% CL) on cosmic-ray photon fraction
derived in the present analysis (Auger) and
previously from AGASA (A1)~\cite{shinozaki}, (A2)~\cite{risse05}
and Haverah Park (HP)~\cite{ave} data compared to some estimates
based on non-acceleration models~\cite{models}.}
\label{fig2}
\end{minipage}
\end{figure}

The statistical method for deriving an upper limit follows that
introduced in~\cite{risse05}.
In brief, for each event a $\chi^2$ value is derived by comparing the observed
$X_{\rm max}$ to the prediction from photon shower simulations.
Accounting for the limited event statistics,
the chance probability $p(f_\gamma)$ is calculated to obtain data sets
with $\chi^2$ values larger than observed as a function of the hypothetical
primary photon fraction $f_\gamma$.
The upper limit $f_\gamma^{\rm ul}$, at a confidence level $\alpha$, is
then obtained from $p(f_\gamma \ge \epsilon f_\gamma^{\rm ul}) \le 1-\alpha$,
where the factor $\epsilon = 0.88$ accounts for the different detector acceptance
to photon and nuclear primaries.

For the Auger data sample, an upper limit on the 
photon fraction of 26\% at a confidence level of 95\% is derived.
In Fig.~2, this upper limit is plotted together with previous experimental
limits and some estimates based on non-acceleration models.
The presented 26\% limit confirms and improves the existing limits 
above $10^{19}$~eV.

\section{Discrimination power of surface array observables}
In 5 out of the 16 selected events, the number of triggered surface
detectors is large enough to perform a standard SD
reconstruction~\cite{bertou05}.
Several observables can be used for primary photon
discrimination~\cite{bertou00}, e.g.:
\begin{itemize}
\item[$\bullet$] {\em rise time:} For each triggered tank, we define a
rise time
as the time for the integrated signal to go from 10\% to 50\% of its
total value.
By interpolation between rise times recorded by the tanks at different
distances to the shower core, the rise time at 1000~m core distance
is extracted after correcting
for azimuthal asymmetries in the shower front.
Compared to nuclear primaries, where the rise time is relatively short due to 
muons that do not suffer from multiple scattering as shower electrons do,
rise times in muon-poor photon showers are expected to be
significantly larger.
\item[$\bullet$] {\em curvature:} The shower front shape is fitted to a
sphere (expanding at speed of light as the shower propagates)
using the start times of the FADC traces of each
station. Then the curvature for the event is defined as the inverse of
the radius of the sphere at the shower core position on ground.
As the photon-initiated showers in general develop deeper in the
atmosphere, the shower front curvature is expected to be larger than
that of nuclear primaries.
\end{itemize}

\begin{figure}[b]
\noindent
\begin{minipage}[l]{.48\linewidth}
\includegraphics[width=.9\textwidth]{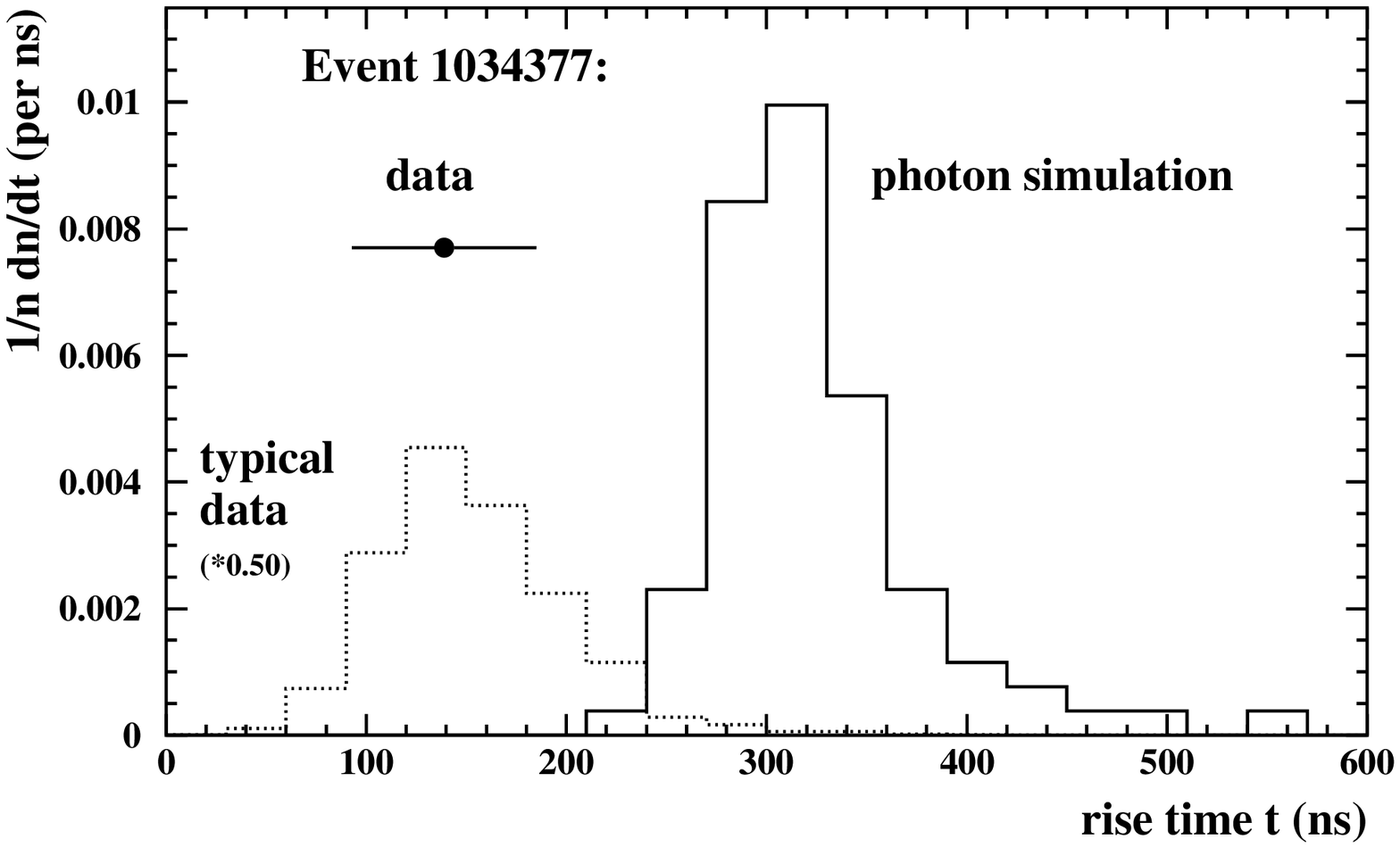}
\caption{Example of rise time measured in an individual shower
(same as in Fig.~\ref{fig1})
(point with error bar) compared to the $X_{\rm max}$ distribution
expected for photon showers (solid line).
The typical data distribution from SD events
at comparable zenith angle is also given (dashed line;
normalization changed as indicated).}
\label{fig3}
\end{minipage}\hfill
\begin{minipage}[c]{.48\linewidth}
\includegraphics[width=.9\textwidth]{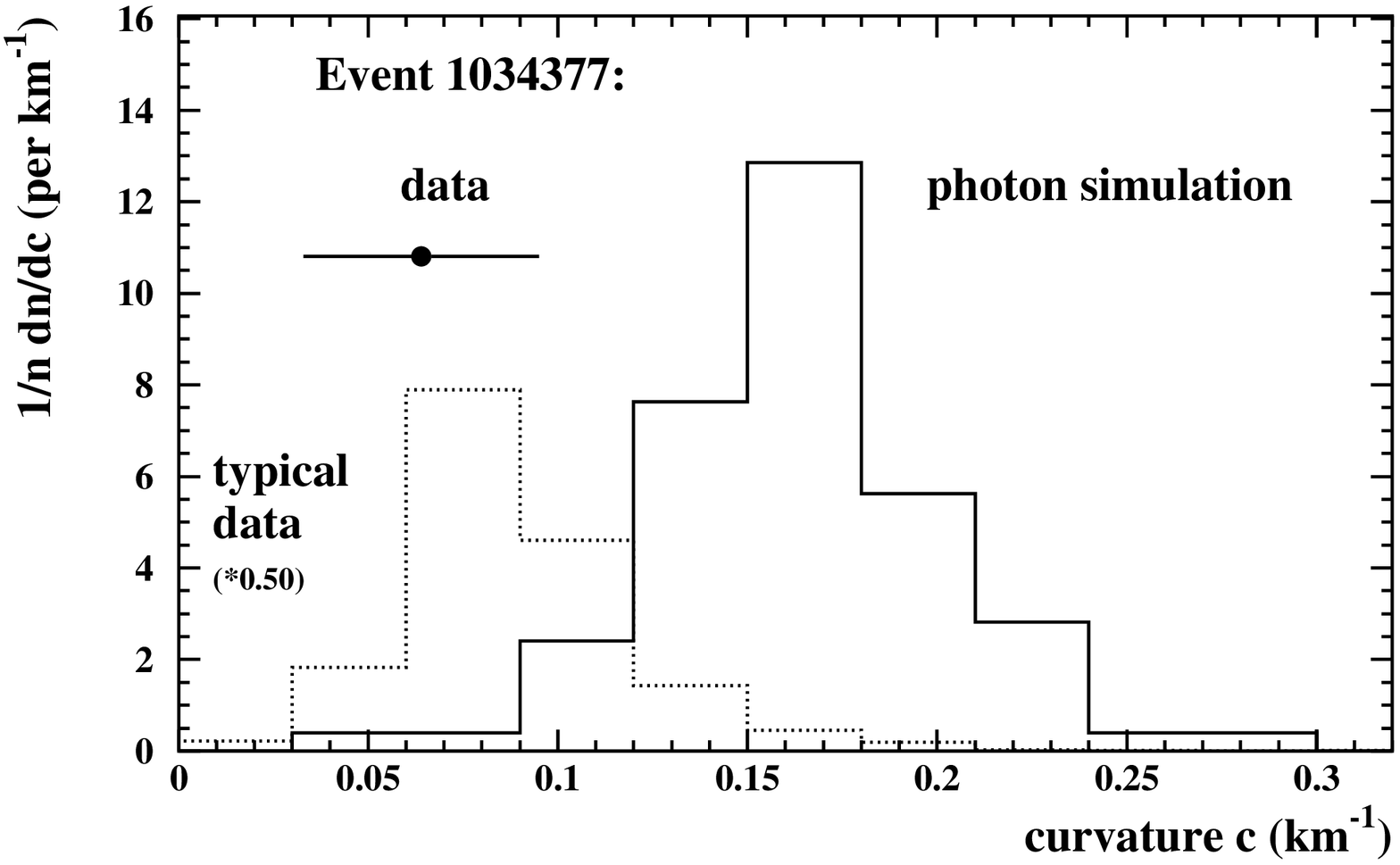}
\caption{Example of curvature measured in an individual shower
(same as in Fig.~\ref{fig1})
(point with error bar) compared to the $X_{\rm max}$ distribution
expected for photon showers (solid line).
The typical data distribution from SD events
at comparable zenith angle is also given (dashed line;
normalization changed as indicated).}
\label{fig4}
\end{minipage}
\end{figure}

As an example, for the specific event shown in Fig.~\ref{fig1},
the measured rise time and curvature data are compared
to the  simulated distributions in Figs.~\ref{fig3} and~\ref{fig4}.
For this and the other SD reconstructed hybrid events,
the SD observables are well separated
from the predictions for primary photons.
These results provide independent information
to the photon limit derived by the hybrid analysis.
They support the conclusion that a photon origin of
the observed events is not favoured.

The SD data statistics at these energies is considerably larger than
the hybrid statistics, as the duty cycle of the
fluorescence telescopes is $\sim$10\%.
To exploit the excellent statistical power,
which will allow us to test hypothetical
primary photon fractions that are significantly smaller,
current studies are performed on subtleties specific to
an SD-only analysis:
(i) event trigger and reconstruction are not fully efficient
for photons at $10^{19}$eV;
(ii) the primary energy estimation is mass dependent, which could
lead to a selection bias.

\section{Outlook}
The photon bound derived in this work is mainly limited by the small 
number of events.
The data statistics of hybrid events will considerably increase in
the near future, and much lower primary photon fractions can be tested.
Moreover, the larger statistics
will allow us to increase the threshold energy of $10^{19}$~eV chosen
in the present analysis to energy ranges where even larger
photon fractions are predicted by some models.

The discrimination power of surface detector observables will be
further exploited.  If hybrid detection is not required then
statistics is significantly increased.
Ways to reduce a possible selection bias in SD-only analyses
are being investigated. Also, the technique introduced in~\cite{ave}
where event rates of near-vertical and inclined showers are compared
to each other, can be further developed.

The uncertainty in extrapolating the photonuclear cross-section
to highest photon energies imposes a systematic uncertainty
in photon shower simulations both for fluorescence light
and ground particle observations~\cite{risse04,risse05}.
Related systematic studies are ongoing.

\end{document}